\documentclass[conference,10pt,a4paper]{IEEEtran}
\usepackage{xcolor}
\usepackage{amsmath}
\usepackage{times}
\usepackage{graphicx}
\usepackage{multirow}
\usepackage[none]{hyphenat}
\usepackage{float}
\usepackage{subfig}
\usepackage{iftex}
\usepackage{quotes} 
\usepackage[backend=bibtex,sorting=none,style=ieee]{biblatex}%
\ifPDFTeX%
\usepackage{t1enc}
\usepackage{times}
\fi%
\ifXeTeX%
\usepackage{fontspec}
\fi%
\ifLuaTeX%
\usepackage{fontspec}
\fi%
\makeatletter

\def\@maketitle{\newpage
\bgroup\par\addvspace{0.5\baselineskip}\centering%
\ifCLASSOPTIONtechnote
   {\bfseries\large\@IEEEcompsoconly{\sffamily}\@title\par}\vskip 1.3em{\lineskip .5em\@IEEEcompsoconly{\sffamily}\@author
   \@IEEEspecialpapernotice\par{\@IEEEcompsoconly{\vskip 1.5em\relax
   \@IEEEtitleabstractindextextbox{\@IEEEtitleabstractindextext}\par
   \hfill\@IEEEcompsocdiamondline\hfill\hbox{}\par}}}\relax
\else
   \vskip0.2em{\EuMWtitlesize\ifCLASSOPTIONtransmag\bfseries\LARGE\fi\@IEEEcompsoconly{\sffamily}\@IEEEcompsocconfonly{\normalfont\normalsize\vskip 2\@IEEEnormalsizeunitybaselineskip
   \bfseries\Large}\@title\par}\vskip1.0em\par
   \ifCLASSOPTIONconference%
      {\@IEEEspecialpapernotice\mbox{}\vskip\@IEEEauthorblockconfadjspace%
       \mbox{}\hfill\begin{@IEEEauthorhalign}\@author\end{@IEEEauthorhalign}\hfill\mbox{}\par}\relax
   \else
      \ifCLASSOPTIONpeerreviewca
         {\@IEEEcompsoconly{\sffamily}\@IEEEspecialpapernotice\mbox{}\vskip\@IEEEauthorblockconfadjspace%
          \mbox{}\hfill\begin{@IEEEauthorhalign}\@author\end{@IEEEauthorhalign}\hfill\mbox{}\par
          {\@IEEEcompsoconly{\vskip 1.5em\relax
           \@IEEEtitleabstractindextextbox{\@IEEEtitleabstractindextext}\par\hfill
           \@IEEEcompsocdiamondline\hfill\hbox{}\par}}}\relax
      \else
         \ifCLASSOPTIONtransmag
           {\@IEEEspecialpapernotice\mbox{}\vskip\@IEEEauthorblockconfadjspace%
            \mbox{}\hfill\begin{@IEEEauthorhalign}\@author\end{@IEEEauthorhalign}\hfill\mbox{}\par
           {\vspace{0.5\baselineskip}\relax\@IEEEtitleabstractindextextbox{\@IEEEtitleabstractindextext}\vspace{-1\baselineskip}\par}}\relax
         \else
           {\lineskip.5em\@IEEEcompsoconly{\sffamily}\sublargesize\@author\@IEEEspecialpapernotice\par
           {\@IEEEcompsoconly{\vskip 1.5em\relax
            \@IEEEtitleabstractindextextbox{\@IEEEtitleabstractindextext}\par\hfill
            \@IEEEcompsocdiamondline\hfill\hbox{}\par}}}\relax
         \fi
      \fi
   \fi
\fi\par\addvspace{0.0\baselineskip}\egroup}

\def\EuMWtitlesize{\@setfontsize{\EuMWtitlesize}{24}{24pt}}
\def\EuMWauthorsize{\@setfontsize{\EuMWauthorsize}{11}{11pt}}
\def\EuMWaffilsize{\@setfontsize{\EuMWaffilsize}{10}{10pt}}
\def\EuMWcaptionsize{\@setfontsize{\EuMWcaptionsize}{9}{10pt}}
\def\EuMWbibsize{\@setfontsize{\EuMWbibsize}{8}{10pt}}

\def\@IEEEauthorblockNstyle{\EuMWauthorsize\@IEEEcompsocnotconfonly{\sffamily}\@IEEEcompsocconfonly{\large}}
\def\@IEEEauthorblockAstyle{\EuMWaffilsize\@IEEEcompsocnotconfonly{\sffamily}\@IEEEcompsocconfonly{\itshape}\@IEEEcompsocconfonly{\large}}
\def\@IEEEauthordefaulttextstyle{\EuMWauthorsize\@IEEEcompsocnotconfonly{\sffamily}\sublargesize}

\def\thebibliography#1{\section*{\refname}%
    \addcontentsline{toc}{section}{\refname}%
    \EuMWbibsize\@IEEEcompsocconfonly{\small}\vskip 0.3\baselineskip plus 0.1\baselineskip minus 0.1\baselineskip
    \list{\@biblabel{\@arabic\c@enumiv}}%
    {\settowidth\labelwidth{\@biblabel{#1}}%
    \leftmargin\labelwidth
    \advance\leftmargin\labelsep\relax
    \itemsep \IEEEbibitemsep\relax
    \usecounter{enumiv}%
    \let\p@enumiv\@empty
    \renewcommand\theenumiv{\@arabic\c@enumiv}}%
    \let\@IEEElatexbibitem\bibitem%
    \def\bibitem{\@IEEEbibitemprefix\@IEEElatexbibitem}%
\def\newblock{\hskip .11em plus .33em minus .07em}%
\ifCLASSOPTIONtechnote\sloppy\clubpenalty4000\widowpenalty4000\interlinepenalty100%
\else\sloppy\clubpenalty4000\widowpenalty4000\interlinepenalty500\fi%
    \sfcode`\.=1000\relax}

\long\def\@makecaption#1#2{%
\ifx\@captype\@IEEEtablestring%
\par\@IEEEtabletopskipstrut
\else
\@IEEEfigurecaptionsepspace
\fi
\setbox\@tempboxa\hbox{\normalfont\footnotesize {#1.}\nobreakspace\nobreakspace #2}%
\ifdim \wd\@tempboxa >\hsize%
\setbox\@tempboxa\hbox{\normalfont\footnotesize {#1.}\nobreakspace\nobreakspace}%
\parbox[t]{\hsize}{\normalfont\footnotesize\noindent\unhbox\@tempboxa#2}%
\else
\ifCLASSOPTIONconference \hbox to\hsize{\normalfont\footnotesize\hfil\box\@tempboxa\hfil}%
\else \hbox to\hsize{\normalfont\footnotesize\box\@tempboxa\hfil}%
\fi\fi
\ifx\@captype\@IEEEtablestring%
\@IEEEtablecaptionsepspace
\else
\fi}

\newlength\tablecaptiontotableskip
\newlength\figuretocaptionskip
\def\@IEEEfigurecaptionsepspace{\vskip\figuretocaptionskip\relax}%
\def\@IEEEtablecaptionsepspace{\vskip\tablecaptiontotableskip\relax}%

\def\abstract{\normalfont%
\@IEEEabskeysecsize\bfseries\textit{\abstractname}\,\bfseries\textit{---}\,%
\@IEEEgobbleleadPARNLSP}%

\def\IEEEkeywords{\normalfont%
\@IEEEabskeysecsize\bfseries\textit{\IEEEkeywordsname}\,\bfseries\textit{---}\,%
\@IEEEgobbleleadPARNLSP}%
\def\endIEEEkeywords{\relax\vspace{0.67ex}%
\par\if@twocolumn\else\endquotation\fi%
\normalsize\normalfont}%

\def\@IEEEauthorblockNtopspace{0ex}
\def\@IEEEauthorblockAtopspace{1mm}
\def\IEEEkeywordsname{Keywords}
\def\subsubsection{\@startsection{subsubsection}{3}{\z@}{1.5ex plus 1.5ex minus 0.5ex}%
{0.7ex plus .5ex minus 0ex}{\normalfont\normalsize\itshape}}%
\newlength{\CPheadmatchindent}%
\def\@seccntformat#1{\hbox to\CPheadmatchindent{\csname the#1dis\endcsname}\hskip 0.1em \relax}
\IEEEilabelindentA \parindent
\IEEEilabelindent \IEEEilabelindentA
\IEEEelabelindent \parindent
\IEEEdlabelindent \parindent
\IEEElabelindent \parindent
\makeatother

\title{Physical Modeling of Saturated Common Mode Choke}

\author{
    \IEEEauthorblockN{%
        Anna Tak\'acs\IEEEauthorrefmark{1}\IEEEauthorrefmark{2},
        Bal\'azs Gy\"ure-Garami\IEEEauthorrefmark{1}\IEEEauthorrefmark{2}, 
        \'Ad\'am Zolt\'an \'Abrah\'am\IEEEauthorrefmark{1},\\
        P\'eter Tam\'as Benkő \IEEEauthorrefmark{1},
        Norbert M. Nemes \IEEEauthorrefmark{3},
        Ferenc Simon \IEEEauthorrefmark{2}\IEEEauthorrefmark{4}\IEEEauthorrefmark{5},
        Bence Bern\'ath\IEEEauthorrefmark{1}\IEEEauthorrefmark{6}
    }
    \IEEEauthorblockA{%
        \IEEEauthorrefmark{1}Robert Bosch Kft., Hungary\\
        \IEEEauthorrefmark{2}Department of Physics, Institute of Physics, Budapest University of Technology and Economics,\\ Műegyetem rkp. 3., H-1111 Budapest, Hungary\\
        \IEEEauthorrefmark{3}GFMC, Departamento de F\'{i}sica de Materiales, Universidad Complutense de Madrid, Madrid 28040, Spain\\
        \IEEEauthorrefmark{4}Institute for Solid State Physics and Optics, HUN-REN Wigner Research Centre for Physics, Hungary\\
        \IEEEauthorrefmark{5}Stavropoulos Center for Complex Quantum Matter, Department of Physics and Astronomy,\\ University of Notre Dame, Notre Dame, Indiana 46556, USA\\
        \IEEEauthorrefmark{6}\texttt{Bence.Bernath@hu.bosch.com}
    }
}

\begin{document}

\maketitle

\begin{abstract}
Common mode chokes (CMCs) are conventional circuit elements performing several tasks, including noise suppression, hindering electromagnetic interference, providing signal integrity, and circuit protection. Much as they are widely used, their fundamental construction and description are often qualitative and lack an understanding of the underlying physical principles. We discuss the behavior of a commercial CMC based on the physical description of the superparamagnetic core and parasitic circuit elements.
The results are validated using a DC bias current and an external magnetic field, which affect the magnetic properties. The behavior of the CMCs in the strongly non-linear regime is also described. 
\end{abstract}

\section{Introduction}

\IEEEPARstart{C}{ommon} mode chokes (CMCs) are critical components in mitigating electromagnetic interference (EMI) in electronic circuits, particularly in power electronics and signal integrity applications. These inductive devices are designed to suppress high-frequency noise and enhance the electromagnetic compatibility (EMC) of electronic systems. The need for effective EMI suppression has grown significantly with the increasing complexity and integration density of modern electronics, where high-speed switching and compact designs often lead to significant common mode noise issues \cite{Ott2009}.

Though the CMCs are critical components, the role of the underlying material properties of the core and the non-linear magnetization effects on their high-frequency behavior are less explored. Existing models usually consider a parallel RLC circuit as an efficient approximation of the CMC but disregard the physical properties of the core material \cite{dominguez-palacios_characterization_2018, ojeda-rodriguez_simple_2022, ojeda-rodriguez_modal_2023}. There are many important cases when the saturation of the core material significantly alters the behavior of the CMC (e.g. electrostatic discharge exposure \cite{ammer_characterizing_2019}, DC superimposition or bias \cite{nomura_high-frequency_2024, imaoka_modeling_2019}, leakage inductance \cite{nemashkalo_unexpected_2023}). Therefore, a purely RLC circuit cannot give fundamental insights as it models the behavior in the linear regime only.

However, even without theoretical understanding, there is a pragmatic approach for the engineering community to decide whether a CMC can perform sufficiently if its core is magnetized. Using a DC-bias tee and an impedance analyzer, one can apply current (of which the amplitude imitates a real-life scenario) where the induced magnetic field magnetizes the core, and the RF response of such a system can be obtained. The usual expectation is that the common mode impedance decreases in such cases, while the differential mode impedance does not change significantly. While such a test gives an indication of the behavior of the CMC, it has limitations in a non-ideal condition. The bias-tee has its own frequency characteristics, the DC current might warm up the sample, and the generated magnetic field is not homogeneous. However, it is not trivial to decide whether such limitations pose a serious uncertainty in the measurement, negatively affecting the design of the final filter circuit. 

Motivated by these, we create an ideal condition for magnetizing the core of the CMC by applying an external homogeneous magnetic field achieved by a high-precision electromagnet. We then meticulously compare the results with the bias-tee-based measurements, ensuring the highest level of accuracy in our description. 

We study a commercial common mode choke using an industry-standard measurement approach based on CISPR~17~\cite{CISPR17} to obtain its frequency-dependent impedance. We show that it can be well modeled with magnetic relaxation phenomena, resulting in a Debye-type relaxation behavior~\cite{topping_c_2018}. The effect of the magnetic field, either due to a DC bias current or an external magnetic field, is studied in detail. 
In this paper, we demonstrate that\begin{itemize}
    \item the common-mode impedance can be modeled on the base of the core material properties even with magnetized core,
    \item the CMC characterization with DC bias tee gives sufficient information for solving basic design problems,
    \item the extreme limit of operation of the CMC can be studied by applying external magnetic field.
\end{itemize}

\section{Measurement}

\subsection{Measurement method}

To investigate the radio frequency response, we utilized both a 2-port and a 4-port vector network analyzer (VNA). Measurements were performed in a frequency range of 10~MHz to 1~GHz with a resolution bandwidth of 10~kHz. Using the VNA, we measured the $S$-parameters of the system under examination, allowing us to calculate the common-mode and differential-mode impedance of the CMC according to CISPR 17.

\begin{figure}[H]
    \centering
    \includegraphics[width=\columnwidth]{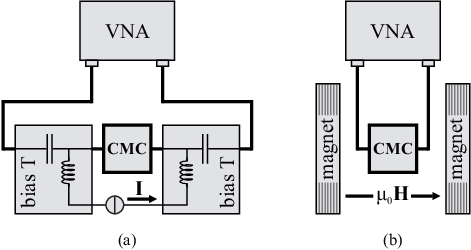}
    \caption{Characterization setups: (a) with DC bias tee; (b) with external magnetic field. The measurement device is a VNA. Note the presence of two DC bias tee circuits and a current source on the left hand side. On the right hand side, an electromagnet generates a homogeneous magnetic field.}
    \label{measurement_setups}
\end{figure}

In tests involving DC bias current, we used two commercial bias tees (see Fig. \ref{measurement_setups}a), which  define the lower frequency limit of our measurements, as they operate from 10~MHz. As we gradually increased the DC current through the coils of the CMC, the RF response was recorded.

For the magnetic field measurements, the CMC was positioned between two disks of a precise laboratory electromagnet (as shown in Fig. \ref{measurement_setups}b).
During the measurement process, we gradually increased the bias current and the magnetic field strength while recording the RF response using the VNA.

Throughout this text, we use the term ‘bias' to refer to the general influence stemming from either the DC bias tee or the external magnetic field.
\begin{figure}[]
\centering
\subfloat[]{
\centering
\includegraphics[width=0.48\columnwidth]{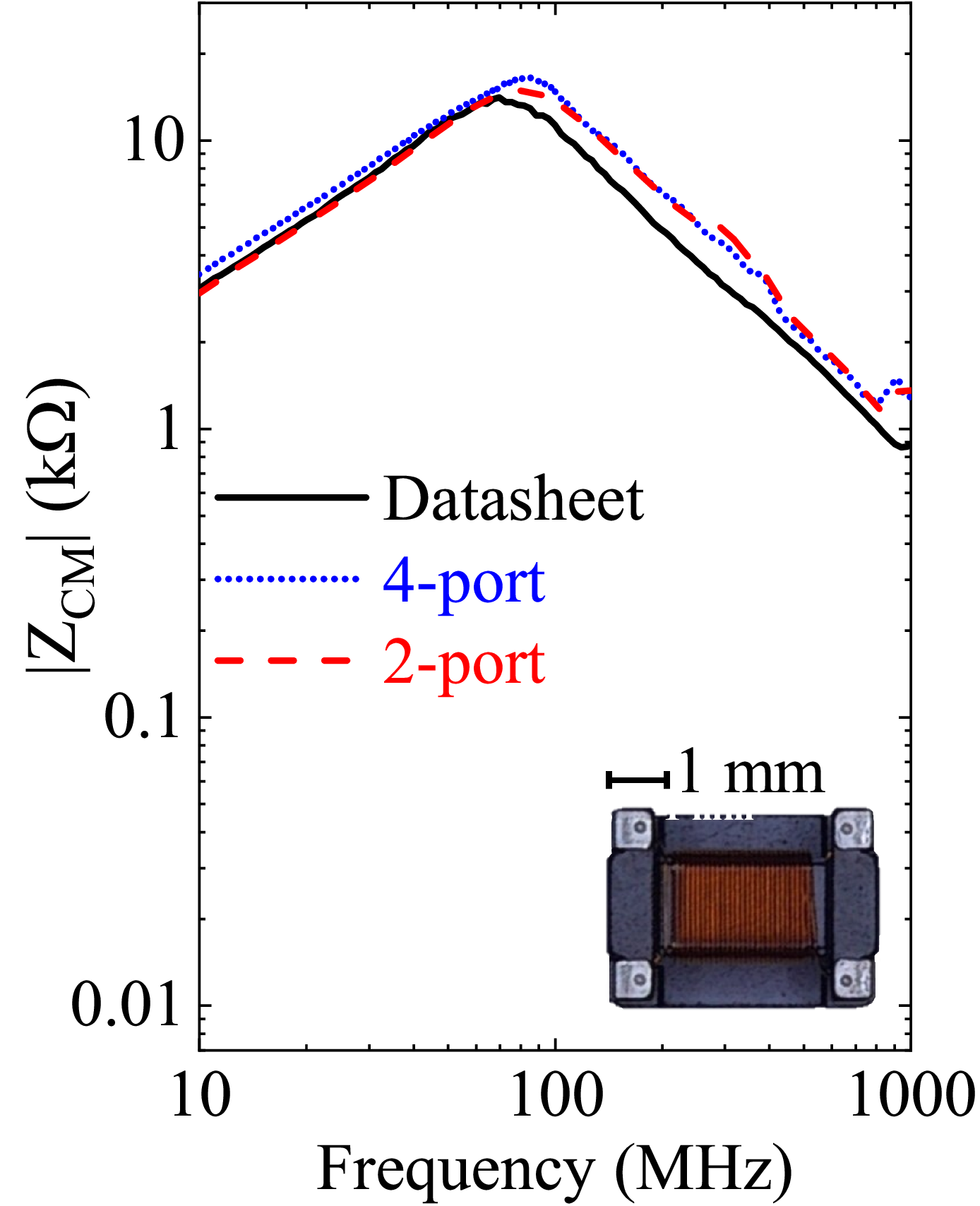}
\label{datasheet_comp}
}
\subfloat[]{
\centering
\includegraphics[width=0.48\columnwidth]{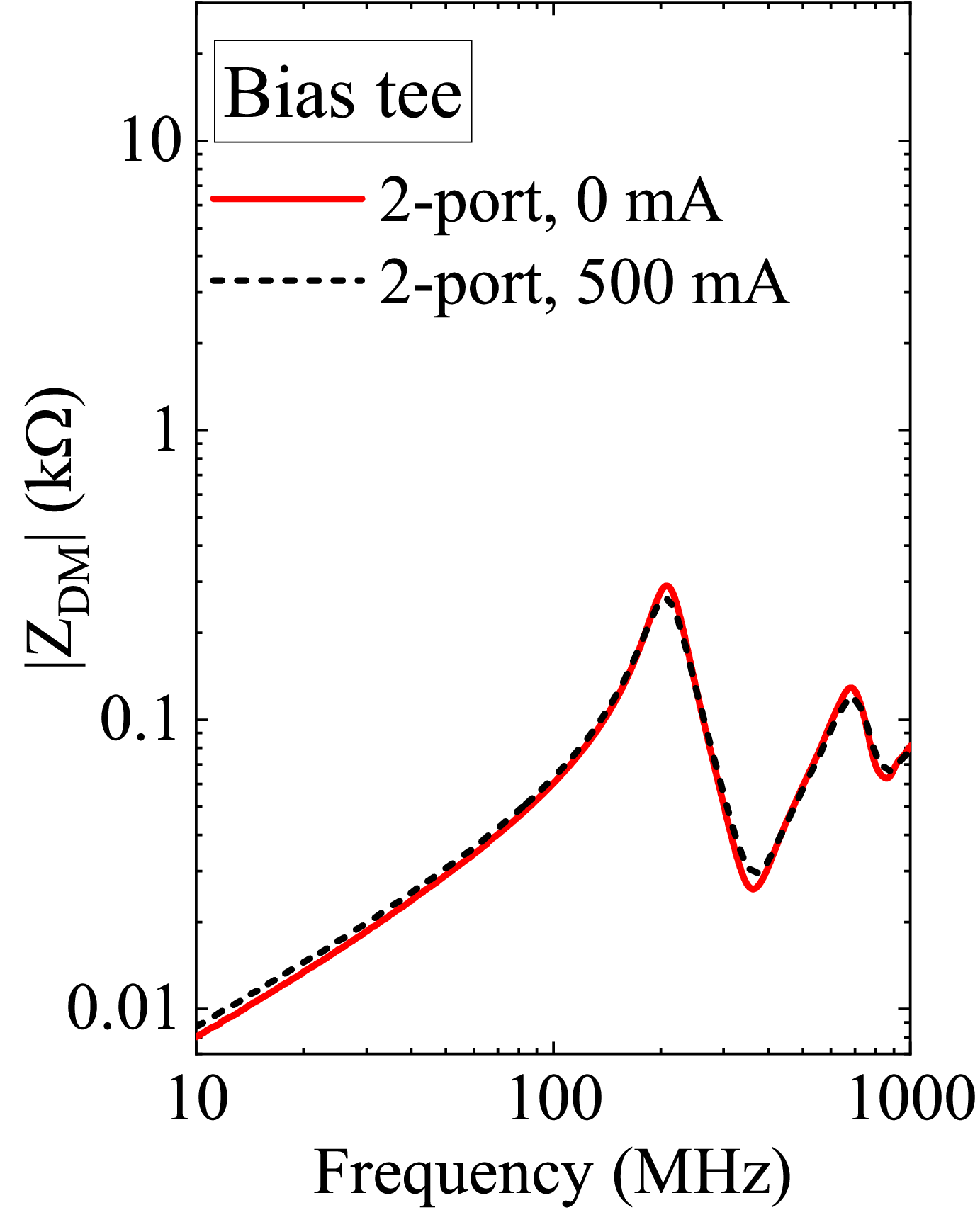}
\label{DM}
}
\\[2.6mm]
\caption{(a) Common mode impedance of the CMC. The solid black line is the manufacturer provided data \cite{act45}, the dotted blue and the dashed red lines are our measurement results with the 4- and 2-port fixtures, respectively. (b) Differential mode impedance of the CMC. The solid red is the result when there is no bias current, the dashed black is the impedance in the presence of 500 mA bias current.}
\label{all_curves}
\vspace{-\baselineskip}
\end{figure}

Fig. \ref{datasheet_comp} presents three common-mode impedance curves. The black line is obtained from the 4-port $S$-parameters in the datasheet \cite{act45}. The blue dotted line represents data obtained from the 4-port $S$-parameter measurement, while the red dashed line reflects the results from the 2-port measurement. In the latter configuration, the two ends of the coils are connected, allowing us to measure two parallel coils as common mode \cite{CISPR17}. From a practical standpoint, our in-house measurements align well with the datasheet. Additionally, the 2-port measurement closely resembles the 4-port test; therefore, we chose to maintain this simpler configuration for tests involving the DC bias tee and external magnetic field.

Fig. \ref{DM} illustrates the differential-mode impedance of the CMC from a 2-port measurement (the coils are in series \cite{CISPR17}). The solid red line indicates the measurement obtained without bias, while the dashed black line represents the results with a 500 mA bias current. Unlike the common-mode response (shown in Fig. \ref{fig:compare1}), the differential-mode response remains unaffected by the bias. Consequently, this study will focus solely on the common-mode impedance.
\subsection{Common mode choke}

The investigated common mode choke is a wide spread, commercially available surface mount device (SMD) typically used on CAN \cite{CAN} communication buses (TDK ACT45B-510-2P-TL003 \cite{act45}). The core material has an ‘H’ shape and the middle part supports the coil windings (see the photo in Fig. \ref{datasheet_comp}). The coils are wound on top of each other, resulting in an excellent coupling coefficient ($k\approx 1$).
The nominal inductance of a single coil (including the ferrite material) is 51 $\mu$H at 100~kHz in the absence of an external magnetic field or bias current. 

\subsection{Test fixture}

The test fixtures were designed according to the guidelines of CISPR 17. We prepared three fixtures: one 4-port device suitable for the mixed-mode $S$ parameter measurement and two 2-port devices for the common-mode and differential-mode measurements. The size of the boards was designed to be the smallest possible to fit into the electromagnet in several orientations and to have minimal impact on the measurement. Surface-mount end-launch SMA connectors were selected with 0.76~mm diameter center connectors as these have better transmission than vertical or through-hole ones and have sufficiently high current carrying capabilities. 
We chose a simple two-layer board with 1.55~mm FR4 as a substrate for the layer stack, which has no significant effect on the measurement below 1~GHz. The traces from the SMA to the component footprint were designed to be 50~$\Omega$ for impedance matching, as we used standard coaxial (SMA) 50~$\Omega$ cables with the VNA. We applied the series-through-method as described in~\cite{CISPR17}.

\section{Physical modeling}

In many cases, it is quite efficient to model common mode chokes with parallel RLC circuits~\cite{ojeda-rodriguez_simple_2022,ojeda-rodriguez_modal_2023,dominguez-palacios_characterization_2018}. 
The common-mode impedance is 
\begin{multline}
\label{z_cmc1}
    Z^{RLC}_{\text{CM}}(\omega ) =\\
     \frac{1}{2}\left[ j\omega C_t +\frac{1}{j(L+M)\omega}+\frac{1}{R_C} +\sum_{r=1} ^{n} \frac{1}{R_r+j\omega L_r} \right]^{-1}
\end{multline}

The factor of $\frac{1}{2}$ indicates that each coil has its own parallel RLC circuits that are also parallel to each other. The $C_t$ indicates the parasitic capacitance of a single coil due to the winding. $L$ is the inductance of a coil, $M$ is the mutual inductance equals to $k\cdot L$, where $k$ is the coupling coefficient. $R_C$ is the damping factor or the ohmic resistance. Additional parameters can be introduced for CMCs with more complicated impedance curves to achieve a better modeling result: $R_r$ és $L_r$ are resistors and inductors parallel to the other components. All parameters are positive real numbers and do not depend on the frequency. Therefore, the properties of the magnetic core (i.e., the complex magnetic permeability) are not explicitly included. 
Though this model works very well for a several CMCs~\cite{ojeda-rodriguez_simple_2022,ojeda-rodriguez_modal_2023,dominguez-palacios_characterization_2018}, it lacks the physical description of the core itself. 

\begin{figure}[t]
    \centering
    \includegraphics[width=0.5\columnwidth]{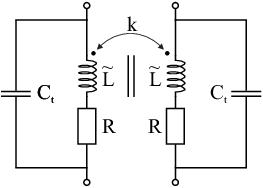}
    \caption{Equivalent circuit diagram of a common mode choke. $C_t$ denotes the parasitic capacitance of a single wire due to the winding, $\tilde{L}$ denotes the inductance of a coil including the ferrite core, $k$ denotes the coupling strength between the two coils. }
    \label{circuit_diagram_of_CMC}
\end{figure}

One can insert the complex susceptibility into $L$ forming a compact model~\cite{Liu, Kelley}  (see Fig. \ref{circuit_diagram_of_CMC}):
\begin{gather}\label{z_cmc2}
   Z^{physical}_{\text{CM}}\left(\omega \right)=  \frac{1}{2} \left[ j\omega C_t +\frac{1}{j\left(1+k\right)\widetilde{L}\omega+R}\right]^{-1} 
\end{gather}
where $R$ is the DC resistance  of a single coil and $\widetilde{L}$ includes the so-called Debye susceptibility~\cite{Kelley,topping_c_2018}, which originates from the description of a driven and damped harmonic oscillator with no inertial terms:
\begin{equation}
    d \dot{x} + Kx = f
\end{equation}
$x$ is the generalized displacement, $\dot{x}$ is its time derivative, $f$ is the driving external force, $d$ is the damping factor, and $K$ is the generalized spring constant. The solution of this differential equation in the time domain is an exponentially decaying function, while it is a Lorentzian function in the frequency domain. Therefore, the frequency dependence of the susceptibility is:
\begin{equation}
    \widetilde{\chi } = \dfrac{\chi _0}{1+j\omega\tau}
\end{equation}
where $\chi _0 = 1/K$ and $\tau = d /K$. If the driving force is a harmonic field which acts on magnetic dipoles, then $\chi _0$ is the static magnetic susceptibility and $\tau$ is the relaxation time of the dipoles.

Therefore, the complex inductivity including the Debye susceptibility reads:

\begin{equation}\label{L_Debye}
    \widetilde{L}=L_0\left(1+\widetilde{\chi}\right)
\end{equation}
where $L_0$ is the air-coil inductivity.

The so-called Havriliak-Negami model\cite{topping_c_2018,Liu}  describes more realistic situations where the relaxation time has a distribution:
\begin{equation}
    \widetilde{\chi}=\chi_S+\frac{\chi_T-\chi_S}{\left[1+\left(j\omega \tau\right)^{1-\alpha}\right]^{\beta}} 
\end{equation}
It also introduces the low and high-frequency limits of the magnetic susceptibility: $\chi _S = \chi (\omega \rightarrow \infty )$ is the adiabatic, $\chi _T = \chi (\omega \rightarrow 0)$ is the isothermal susceptibility. These describe that the adiabatic susceptibility never reaches zero, as the high-frequency (THz, PHz) material properties have some influence on the baseline of the microwave properties. The $\alpha$ and $\beta$ parameters are between 0 and 1 and express how wide ($\alpha$) and symmetric ($\beta$) the susceptibility curve is near $\omega = 1/\tau$. 

Eq. \eqref{L_Debye} can be modified to account for a finite filling factor or demagnetization factors. In practice, this means that with this model, it is not possible to determine the \textit{relative} permeability or \textit{absolute} susceptibility of the core. Therefore, we use the term \textit{effective} when discussing the physical properties of the core. 

The Havriliak-Negami model can be approximated by a series of exponential relaxation (i.e., Debye function) \cite{Kelley}:
\begin{gather}\label{L_Debye_series}
    \chi_S+\frac{\chi_T-\chi_S}{\left[1+\left(j\omega \tau\right)^{1-\alpha}\right]^{\beta}} \approx \chi_S+(\chi_T-\chi_S)\sum_{r=1}^{N_r}\frac{a_r}{1+j\omega\tau_r}
\end{gather}
where $a_r$ is a dimensionless weight, $\tau_r$ is the relaxation time of the $r$th function and $N_r$ is the total number of functions.  It is also known that the simplest circuit network to model an exponential relaxation is built by ohmic and inductive components~\cite{Liu}.
To illustrate this, we compare the simple Havriliak-Negami model ($\alpha=0$ and $\beta=1$) with its equivalent circuit in Table~\ref{table_RL_Debye}.

\begin{table}[]
\caption{Comparison between the Havriliak-Negami model (with $\alpha=0$ and $\beta=1$) and its equivalent circuit.}
\centering
\captionsetup{justification=centering}
\small
\resizebox{\columnwidth}{!}{
\renewcommand{\arraystretch}{1}
\huge
\begin{tabular}{|c|c|}
\hline
 & \\
\textbf{Model} & \textbf{Equivalent circuit}  \\
 & \\
\hline
 & \\
\includegraphics[width=0.32\textwidth]{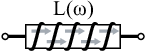} & \includegraphics[width=0.34\textwidth]{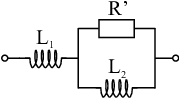}\\ 
 &\\
  & \\
\hline
 & \\
 $Z=j \omega\left(L_0\cdot \left(1 + \chi _S\right) + \dfrac{L_0\cdot \left(\chi _T-\chi _S\right)}{1+j\omega \tau }\right)$ 
 & $Z = j\omega \left(L_1+\dfrac{L_2}{1+j\omega \dfrac{L_2}{R'}} \right)$  \\
  & \\
 \hline \hline
\multicolumn{2}{|c|}{\rule{0pt}{2.5ex}\textbf{Equivalent parameters}} \\
\hline
\multicolumn{2}{|c|}{\rule{0pt}{3ex} $L_1 = L_0\cdot \left(1+\chi _S\right)$} \\
\multicolumn{2}{|c|}{\rule{0pt}{3ex} $L_2 = L_0\cdot \left(\chi _T-\chi _S\right)$}\\
\multicolumn{2}{|c|}{\rule{0pt}{3ex} $L_2/R' = \tau$}\\ 
 \hline
 \end{tabular}
}

\label{table_RL_Debye}
\end{table}

Based on Table \ref{table_RL_Debye}, it is easy to see that we obtain the same equations mathematically for both cases. Also, it is discussed in~\cite{Liu} that adding more stages of $RL$ parallel circuits is equivalent to increasing the number of Debye functions in~(\ref{L_Debye}). Therefore, it is not surprising that the models discussed in \cite{ojeda-rodriguez_simple_2022,ojeda-rodriguez_modal_2023,dominguez-palacios_characterization_2018} work very well as they still capture frequency dependence of the magnetic core, due to the mathematical equivalence of the underlying equations. 

However, there is an advantage to using the Debye picture for modeling a CMC or other inductive elements filled with a magnetic core in case of the external magnetic field with saturation-related effects. All the parameters ($\chi_T$, $\chi_S$, $\tau$) have physical meanings one can follow. The susceptibility values describe the slope of the magnetization curve of the core. In case of increasing external magnetic field, they must decrease. The relaxation time has a very non-linear magnetic field dependence; however, $\tau$ still decreases~\cite{Deissler}, similarly to the susceptibility values.

It is safe to assume that neither $C_t$ nor $R$  depend on the magnetic field, and there is no unexpected magnetic interaction within the PCB that holds the CMC. Therefore, we use \eqref{z_cmc2} to model our datasets.

\section{Discussion}

\begin{figure}[]
    \centering
    \captionsetup{justification=centering}
    \includegraphics[width=0.45\textwidth]{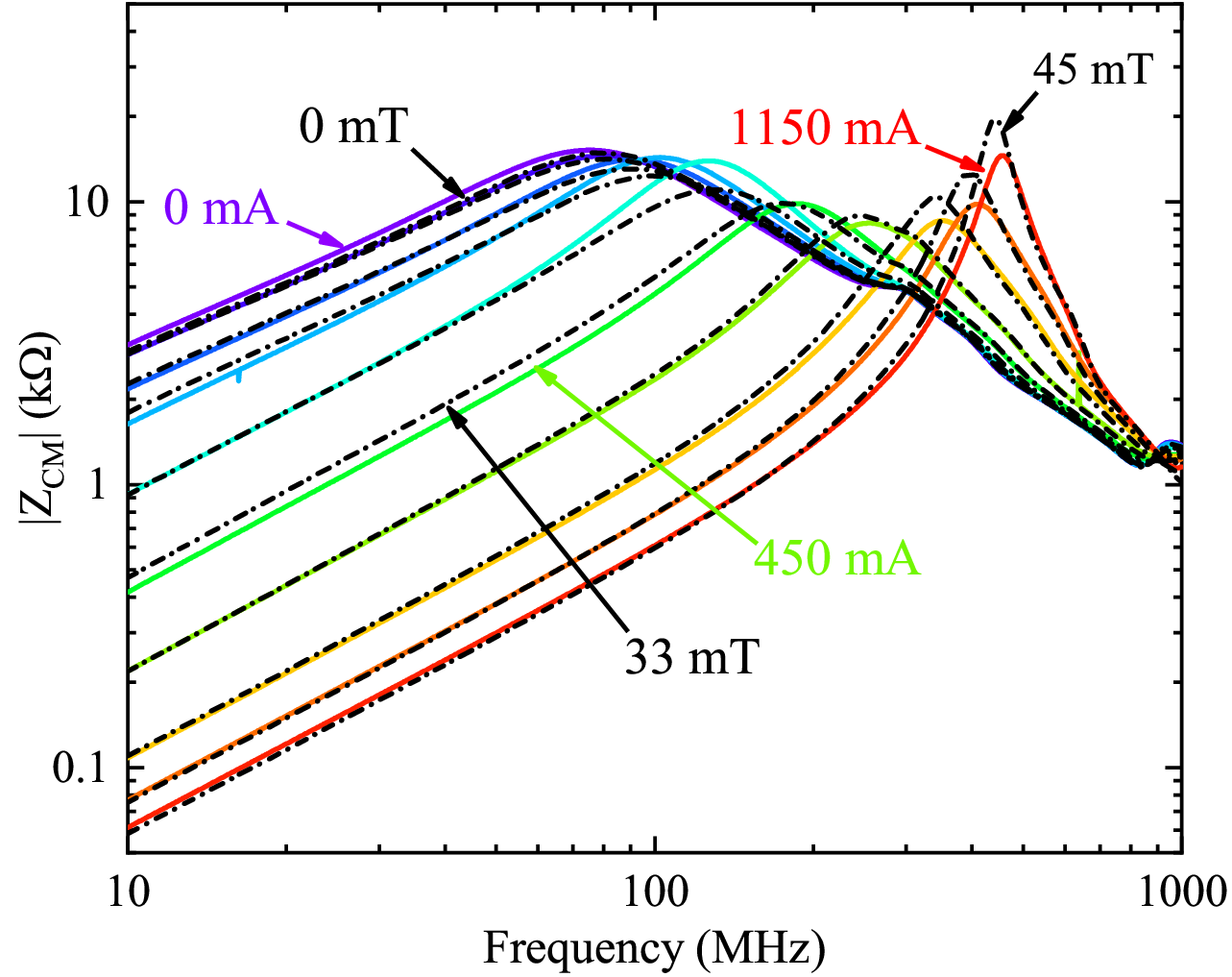}
    \caption{Comparison of the magnetic field and bias current based measurements. Solid colored lines represent the magnetic field measurements, dashed black lines represent the bias current measurements.}
    \label{fig:compare1}
\end{figure}

We selected a few curves from both bias tee (rainbow curves) and external magnetic field (black dash-dot curves) measurements in Fig. \ref{fig:compare1}.  The selection was based  on the similarities regarding the inductive part of the impedance. We invariably observe a resonance peak 
that shifts towards  higher frequencies due to the decreasing inductive part and the constant (parasitic) capacitive part \mbox{($\omega_{\text{res.}}\approx \frac{1}{\sqrt{LC}}$)}. As the magnetic field increases, we observe differences between the two measurements; however, practically, both follow the same trend. We observed a temperature increase during the large current bias measurement (near 1~A, $\sim70^\text{o}$C). However, the observed curves overlap well  with the external magnetic field measurement, which was performed at a constant temperature ($\sim25^\text{o}$C). This means that even with extreme current bias, the dominant effect on the CMC impedance is the magnetization and saturation of the core material. The temperature does not play an important role. Note that the rated current of the CMC is 200~mA according to the datasheet. Therefore, running a test above~1~A  is an overtest, far from the practical design cases, however it helps us map the limitations of our characterization approach.

Both measurements present a somewhat counterintuitive result: as the current or magnetic field increases, the peak impedance values change in a nonmonotonous way, arriving at even higher values than the original unbiased curves, forming a narrow-band filter.
We studied the dependence of the peak height on the parameters of our description and we found that it is closely related to the decrease of the $\tau$ parameter. This is thus considered as a fitting parameter in the following. 

Fig. \ref{all_curves} shows every measured curve with both methods (rainbow curves) and their fit curves from the model (dashed black line). The external magnetic field measurements yield sharper curves towards higher fields. Since the external magnetic field is more homogeneous than the induced magnetic field inside the CMC (bias tee case), we believe the sharpening/broadening effect stems from this field homogeneity issue. However, the model undoubtedly captures these features, as it is explained in the next paragraphs.

We used (\ref{z_cmc2}) as the basis of our fitting process. 

In the first round, the following five parameters determine the impedance curves: $L_0$, the inductance of a single coil without a magnetic core; $ C_t $, the parasitic capacitance resulting from the winding; and the susceptibility values $ \chi_T $ and $ \chi_S $, along with the relaxation time $ \tau $. The role of $\alpha$ and $\beta$ will be discussed later.  The value of $ L_0 $ is derived from purely geometrical factors and the number of turns. We assumed a perfect cylindrical shape, resulting in an estimated inductance of 0.5~µH. However, it is important to note that this is a rough approximation, which increases the uncertainty of the susceptibility values. From a practical standpoint, though, this approximation does not significantly affect the quality of the fit.  Note that an ideal coupling between the two coils was considered, therefore $1+k=2$.

Since $ C_t $ arises from the winding, we assume that this value remains constant during variations in current or magnetic field bias, which is a reasonable assumption. We determined $ C_t $ by performing magnetic field measurements, fitting $ C_t $, $ \chi_T $, $ \chi_S $, and $ \tau $ at both 0~mT and 500~mT. These fits showed that $ C_t $ hardly changes with the bias therefore we took the average value of $ C_t $ from these extreme cases, yielding $ C_t = 0.0721 $~pF. This value was then fixed for all subsequent curves, allowing us to fit the magnetic field (or bias current) dependence of the remaining three parameters: $ \chi_T $, $ \chi_S $, and $ \tau $.

\begin{figure}[]
\centering
\subfloat[]{
\centering
\includegraphics[width=0.48\columnwidth]{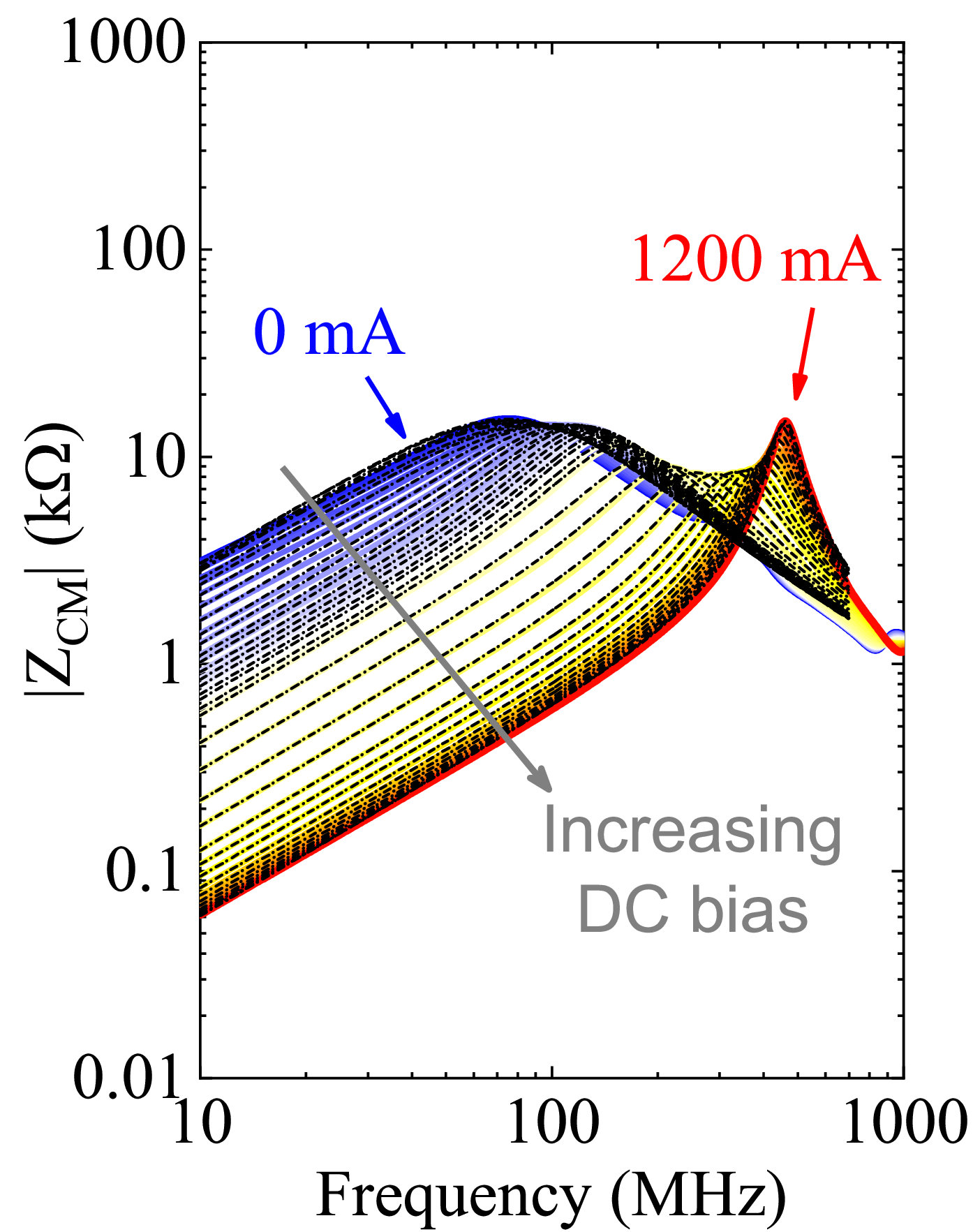}
\label{bias_withfit}
}
\subfloat[]{
\centering
\includegraphics[width=0.48\columnwidth]{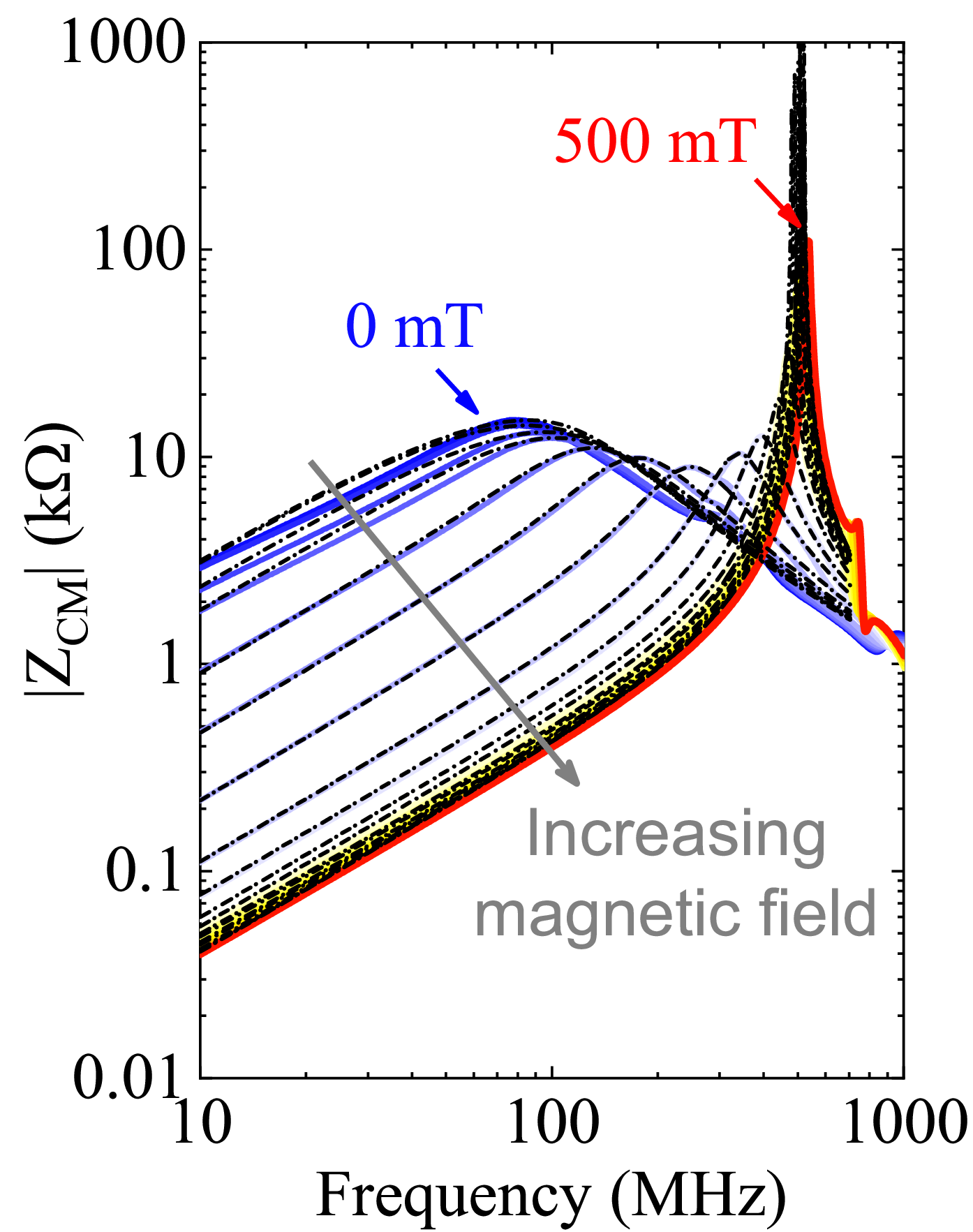}
\label{magn_withfit}
}
\\[2.6mm]
\caption{Fitted curves and the measured data. The solid colored lines represent the measurements, while the dashed black lines show the fitted curves. (a) Measurements under DC bias current. (b) Measurements in an external magnetic field.}
\label{all_curves}
\vspace{-\baselineskip}
\end{figure}

\begin{figure}[t]
\centering
\subfloat[]{
\centering
\includegraphics[width=0.48\columnwidth]{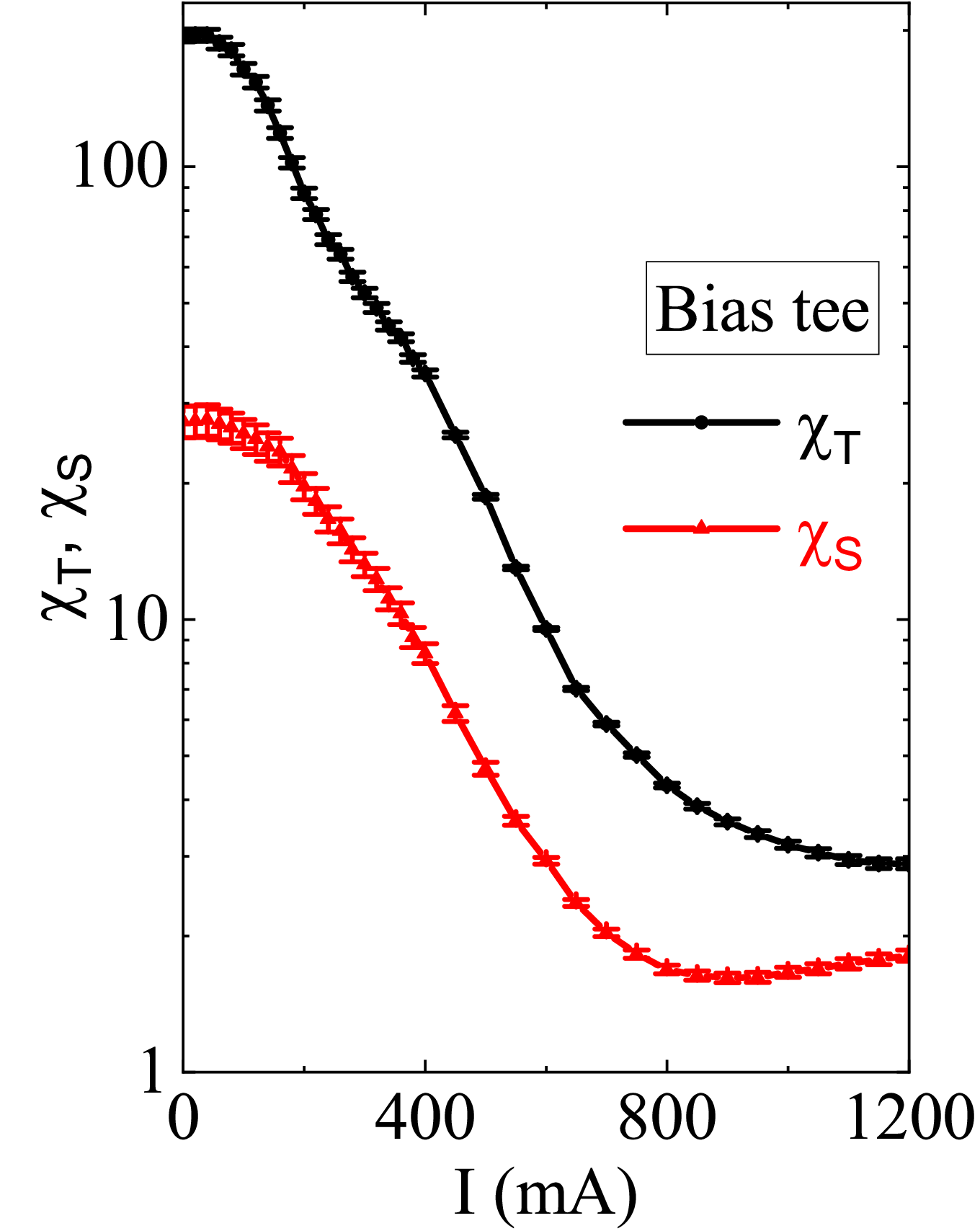}
\label{chi_bias}
}
\subfloat[]{
\centering
\includegraphics[width=0.46\columnwidth]{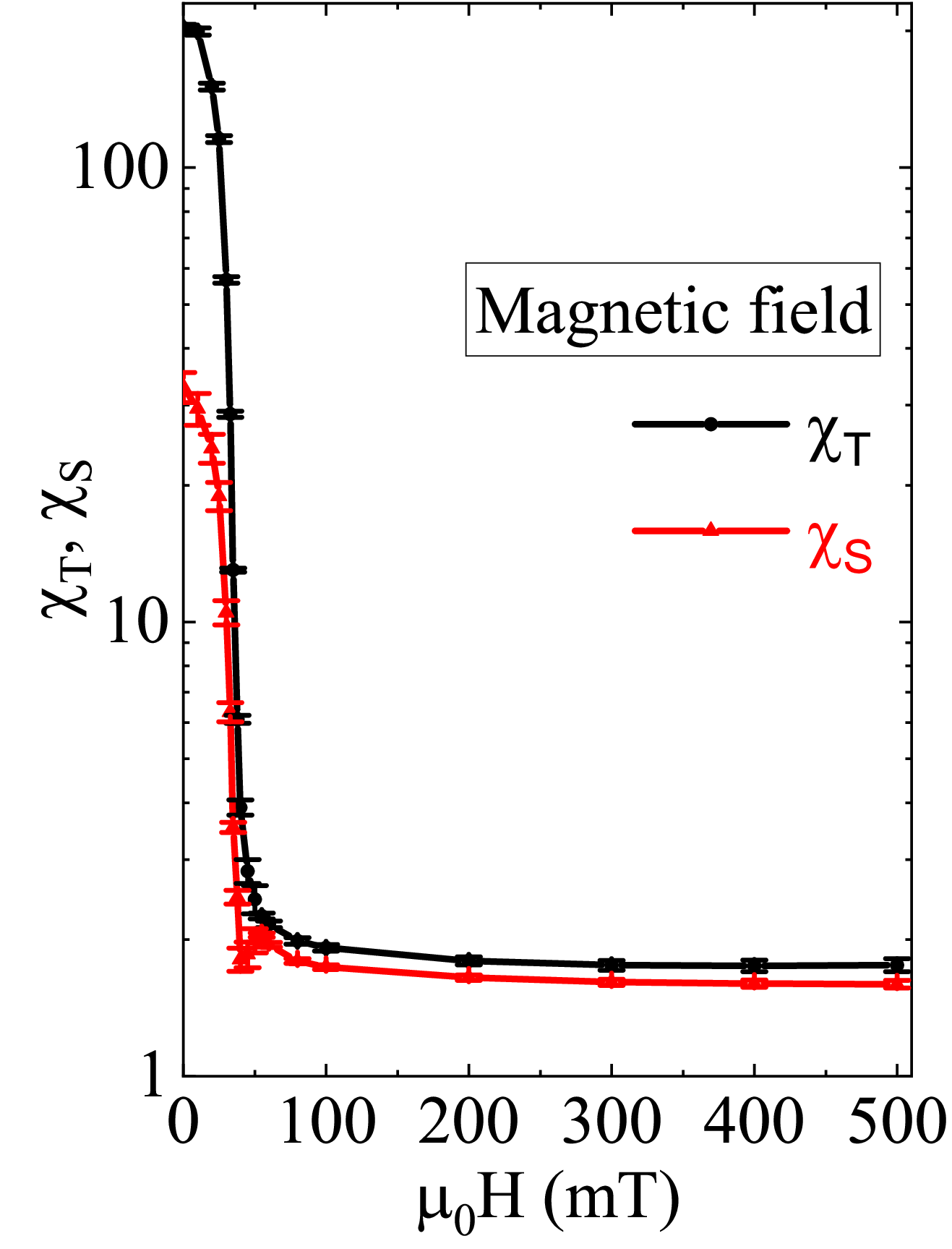}
\label{chi_magn}
}

\vspace{0.5mm}

\subfloat[]{
\centering
\includegraphics[width=0.48\columnwidth]{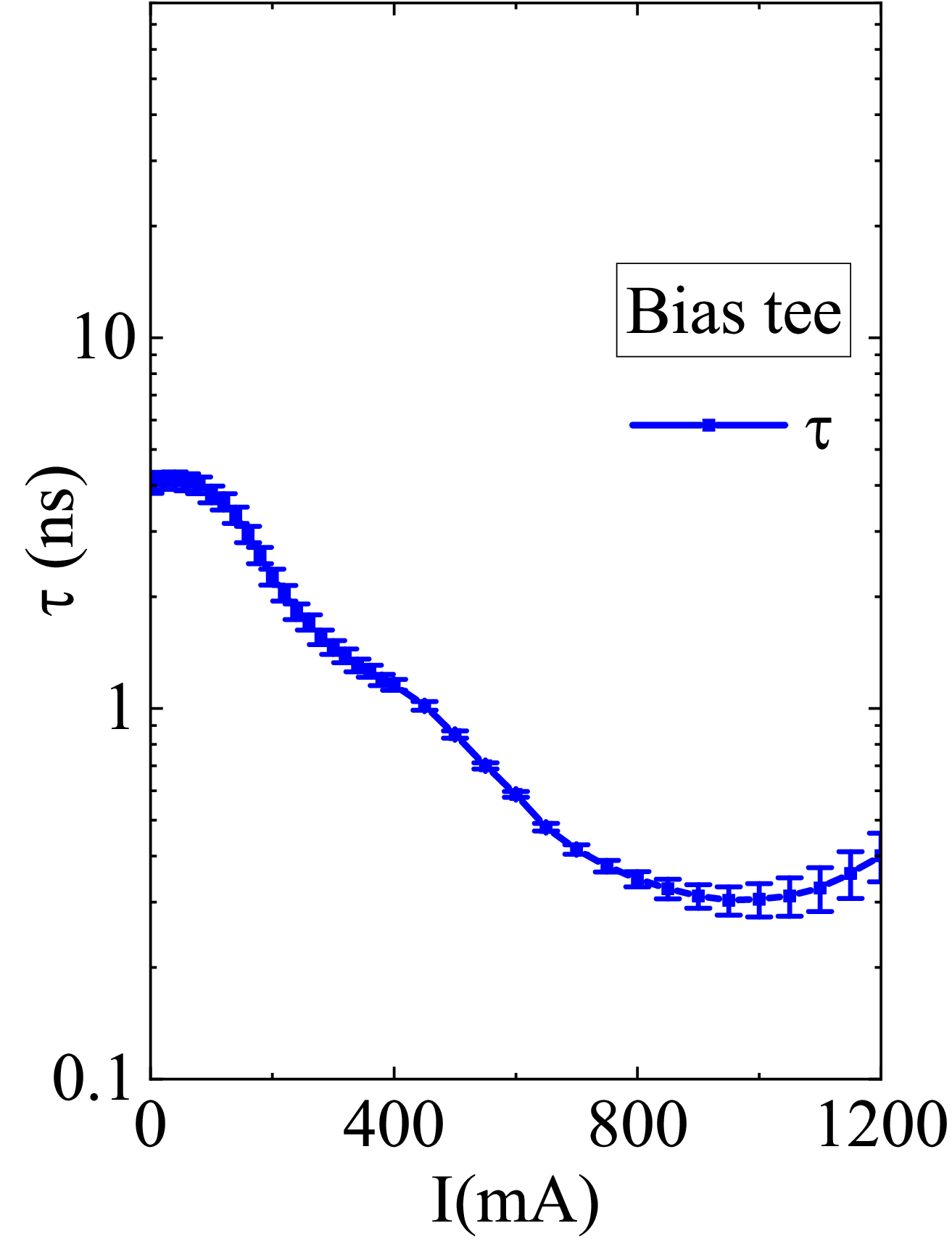}
\label{tau_bias}
}
\subfloat[]{
\centering
\includegraphics[width=0.46\columnwidth]{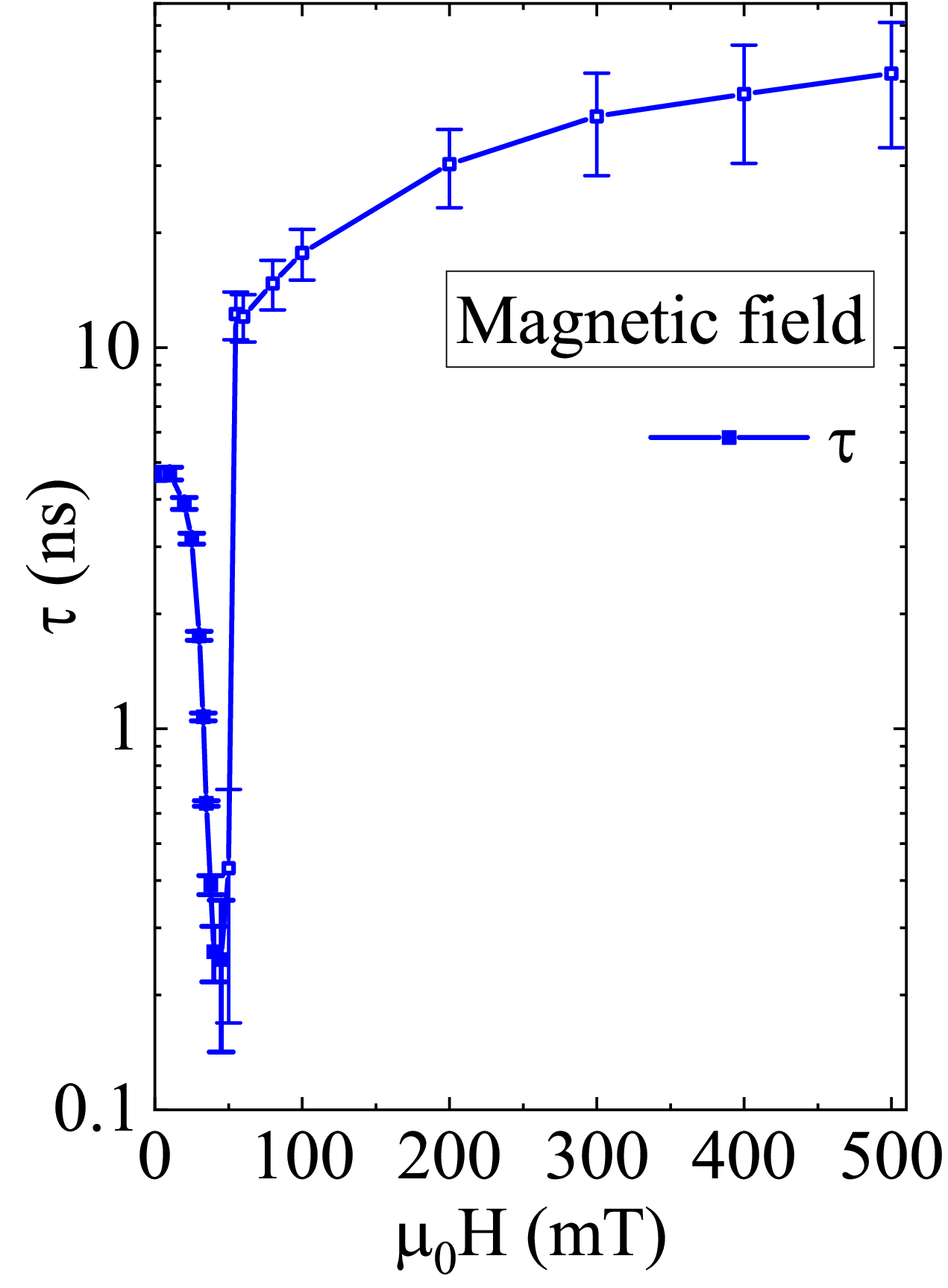}
\label{tau_magn}
}
\\[2.6mm]
\caption{The fitted parameters as the function of the (a) (c) bias current and the (b) (d) magnetic field. The error bars indicate the two-sigma confidence level (95\%). The lines between the points are guides for the eye. On (d) the open-symbol data points are less reliable due to increased uncertainty.}
\label{all_parameters}
\vspace{-\baselineskip}
\end{figure}

To minimize the squared difference between the measurements and the model impedance, we employed the Nelder-Mead simplex algorithm in Python \cite{scikit_rf}. Due to logarithmic sampling and the relatively sharp peaks in the impedance curves, the inductive component of the impedance carries greater weight in the algorithm, which sometimes leads to less accurate representations of the peak and capacitive parts. To address this, we implemented a weighting factor ($ w_{\text{peak}} = 100 $) on the peak position of each curve. The standard error of the fit parameters was also calculated based on~\cite{NumericalRecipes, NelderMead1965}.

Regarding the parameters $\alpha$~and~$\beta$, fitting them simultaneously with the other three parameters is not feasible due to their mutual dependency. Therefore, we fixed $ \alpha $ at 0 and~$\beta $ at 1, allowing us to fit the other parameters. After completing this routine, we varied $ \alpha $ in increments of 0.02 up to 1 and re-fitted the model each time. We calculated the adjusted R-squared (\cite{NumericalRecipes, NelderMead1965}) value for each fit and considered the sum of these values as an indicator of the overall fit quality. After that we fixed $\alpha$ at 0 (best agreement) and started to decrease  $ \beta$ from 1 with the same 0.02 steps. We concluded that the optimal values for our specific case were $ \alpha = 0 $ and $ \beta = 1 $, resulting in the complex inductivity of (\ref{L_Debye}) with only one Debye function. Note that finding the proper $\alpha$ and $\beta$ is  not easy; one needs to tailor the fitting algorithm based on the visual and practical agreement between the data and the model. 

Our model captures very well the inductive part, the peak broadening, and the capacitive part of the impedance curves except for the largest external magnetic fields where the model consistently overshot (up to 1~M$\Omega$) the measurement results. However, our model does not consider the finite resolution bandwidth of the measurement; our model samples with an infinitely narrow frequency window for simplicity.  Furthermore, there is a practical limit to the maximum impedance that a VNA can measure~\cite{practical}, which is not more than 100 k$\Omega$. Therefore, the overshoot issues do not pose serious problems in our analysis.

Fig. \ref{all_parameters} summarizes the DC bias and magnetic field dependence of the three main parameters. In Fig.~\ref{chi_bias} and Fig.~\ref{chi_magn}, we observe monotonously decreasing susceptibility curves plateauing on small but non-zero values. This entirely agrees with the physical intuition; the susceptibility values express the slope of the magnetization curve, regardless of the measurement frequency ($\chi_T$ is the DC limit and $\chi_S$ is the high-frequency limit). As the core experiences an increasing magnetic field, the magnetization starts saturating and flattening, resulting in a smaller derivative value at each field point. Note that the core material was considered as superparamagnetic due to the lack of remanent magnetization, the relatively large susceptibility values and the small saturation magnetic fields \cite{Cullity2008}.

Fig.~\ref{tau_bias} and Fig.~\ref{tau_magn} shows the evolution of the $\tau$ parameter for both bias cases. $\tau$ decreases with the increasing magnetic field strength, agreeing with the theoretical expectations from~\cite{Deissler}, arriving at a minimum point (45~mT and 950~mA). Then, there is an increasing trend with substantially increased error bars. After the minimum of $\tau$,  numerically and physically it is difficult to give reasonable meaning to $\tau$ due to the overall small complex susceptibility values in the saturated region. 

In principle, a core is considered saturated when the external magnetic field no longer increases the magnetization of the material \cite{Griffiths_eldin}. However, due to (super)paramagnetic behavior, the magnetization continues to increase, albeit at a much slower rate after reaching a certain threshold field value.

\renewcommand{\arraystretch}{2} 

\begin{table}[]
\caption{The table summarizes the common-mode impedance drop and the common-mode insertion loss (IL) at 10 MHz under different magnetic field and DC bias current conditions. Z$_{\text{unbiased}}$ is the impedance without external magnetic field or bias current, $\Delta $Z is the Z$_{\text{unbiased}}$ - Z$_{\text{biased}}$ difference at given field or bias current values. Row 1 indicates the impedance and insertion loss values when there is no magnetic field or bias current. Row 2 shows the magnetic field and DC bias current values that cause the impedance to drop by 30\% of the initial impedance. Row 3 indicates the decrease observed at the rated current of the CMC (200~mA). Using interpolation*, we also determined the magnetic field value corresponding to the same decrease in impedance (24 mT). Row 4 shows the decrease observed, when the parameter $\tau $ has its minimum.}
\centering
\small
\begin{tabular}{|c|c|c|c|}
\hline
  & \textbf{Bias curr. / Magn. field} & $\mathbf{Z_{\text{unbiased}}~(\Omega )}$ & \textbf{IL (dB)}\\ \hline
1 & \begin{tabular}[c]{@{}c@{}}Unbiased:\\ 0~mA / 0~mT\end{tabular} & 3040 & 30 \\ \hline \hline
  & \textbf{Bias curr. / Magn. field} & $\frac{|\Delta\mathbf{Z}|}{|\mathbf{Z_{\text{unbiased}}}|}$ & \textbf{IL (dB)} \\ \hline
2 & \begin{tabular}[c]{@{}c@{}}30\% ‘rule'~\cite{nemashkalo_unexpected_2023}:\\ 140~mA / 20~mT\end{tabular} & 30$\% $ & 27 \\ \hline
3 & \begin{tabular}[c]{@{}c@{}}Rated current \cite{act45}:\\  200~mA / 24~mT*\end{tabular} & 55$\% $ & 23 \\ \hline
4 & \begin{tabular}[c]{@{}c@{}}Minimum of the parameter $\tau $:\\ 950~mA / 45~mT\end{tabular} & 98$\% $ & 2 \\ \hline
\end{tabular}
\label{performance_comp}
\end{table}

From an engineering perspective, understanding saturation is crucial for assessing the efficacy of common-mode suppression at specific bias currents or fields. In Table~\ref{performance_comp}, we have summarized our findings at 10 MHz, where inductive behavior is dominant. We compared the change in common-mode impedance and insertion loss at three different bias values. 

Our analysis emphasizes the impedance drop relative to the unbiased reference value (3040 $\Omega$) and the insertion losses (30 dB for the unbiased condition), which are presented in the first row of the table. The second row refers to the 30$\%$~‘rule', derived from a reference work \cite{nemashkalo_unexpected_2023}, which indicates a practical limit for saturation: the inductive impedance drops by 30$\%$ from its unbiased state. The third row presents the effects of the rated current (200~mA, as stated in the datasheet~\cite{act45}). The fourth row illustrates an extreme saturation limit according to our model, where the bias values correspond to the minimum of $\tau$. 
In this scenario, we observe a dramatic case of a 98$\%$ impedance drop, exceeding conventional design expectations. However, at approximately 500~MHz, we achieve a narrow band filter that outperforms the unbiased impedance. Such increased impedance near extreme saturation is the consequence of the  ratio between the imaginary and real part of the inductivity $\widetilde{L}$, originating from the Debye susceptibility.

\section{ Conclusions}
In this study, we compared two techniques for examining the effects of saturated magnetic cores on a signal line common mode choke. We found that using a DC bias tee provides a practical solution for characterizing common mode chokes in most cases. Additionally, applying an external magnetic field allows a better understanding of the pure saturation effect without hindering the characterization process.

We discovered that both current and field bias effects can be effectively modeled based on Debye susceptibility. Our approach is quite general, relying only on real physical phenomena: the superparamagnetic behavior of the core and the parasitic capacitance of the coil resulting from its windings. Therefore, this model could be extended to other inductors, such as power line common mode chokes or simple chokes with magnetic cores.

An exciting avenue for future research would be to investigate the saturation effect on chokes with an external magnetic field oriented in arbitrary directions. Such measurements could reveal potential weaknesses in chokes, ultimately helping to prevent errors in filter designs.

\section*{Acknowledgements}
We acknowledge the National Research, Development and Innovation Office of Hungary (NKFIH) Grants Nr. K137852, K149457, TKP2021-EGA-02, and TKP2021-NVA-02, and 2022-2.1.1-NL-2022-00004. The Spanish Ministry of Science and Innovation is acknowledged for granting the project TED2021-129254-B-C21.

\printbibliography

\end{document}